\documentclass[aps,prl,superscriptaddress,showpacs,floatfix,twocolumn]{revtex4}

\usepackage{graphicx}	

\def\CuCu{Cu+Cu}
\def\dAu{$d$+Au}
\def\AuAu{Au+Au}
\def\pp{$p+p$}
\def\pA{$p+A$}

\def\PbPb{Pb+Pb}
\def\InIn{In+In}

\def\pT{\mbox{$p_{\rm T}$}}

\def\sqrtsNN{\mbox{$\sqrt{s_{NN}}$}}

\def\tab#1{{Table~\ref{#1}}}
\def\fig#1{{Fig.~\ref{#1}}}

\newcommand{\mean}[1]{\left\langle #1 \right\rangle}

\newcommand{\jpsi}{\mbox{$J/\psi$}}
\newcommand{\RAA}{\mbox{$R_{\rm AA}$}}

\newcommand{\Npart}{$N_{\rm part}$}
\newcommand{\Ncoll}{$N_{\rm coll}$}

\newcommand{\raa}{R_{\rm AA}}

\begin{document}

\title{\jpsi~Production in \sqrtsNN~=~200~GeV \CuCu~Collisions}

\newcommand{\abilene}{Abilene Christian University, Abilene, TX 79699, USA}
\newcommand{\banaras}{Department of Physics, Banaras Hindu University, Varanasi 221005, India}
\newcommand{\bnl}{Brookhaven National Laboratory, Upton, NY 11973-5000, USA}
\newcommand{\caucr}{University of California - Riverside, Riverside, CA 92521, USA}
\newcommand{\charlesczech}{Charles University, Ovocn\'{y} trh 5, Praha 1, 116 36, Prague, Czech Republic}
\newcommand{\ciae}{China Institute of Atomic Energy (CIAE), Beijing, People's Republic of China}
\newcommand{\cns}{Center for Nuclear Study, Graduate School of Science, University of Tokyo, 7-3-1 Hongo, Bunkyo, Tokyo 113-0033, Japan}
\newcommand{\colorado}{University of Colorado, Boulder, CO 80309, USA}
\newcommand{\columbia}{Columbia University, New York, NY 10027 and Nevis Laboratories, Irvington, NY 10533, USA}
\newcommand{\czechtech}{Czech Technical University, Zikova 4, 166 36 Prague 6, Czech Republic}
\newcommand{\dapnia}{Dapnia, CEA Saclay, F-91191, Gif-sur-Yvette, France}
\newcommand{\debrecen}{Debrecen University, H-4010 Debrecen, Egyetem t{\'e}r 1, Hungary}
\newcommand{\elte}{ELTE, E{\"o}tv{\"o}s Lor{\'a}nd University, H - 1117 Budapest, P{\'a}zm{\'a}ny P. s. 1/A, Hungary}
\newcommand{\fit}{Florida Institute of Technology, Melbourne, FL 32901, USA}
\newcommand{\fsu}{Florida State University, Tallahassee, FL 32306, USA}
\newcommand{\gsu}{Georgia State University, Atlanta, GA 30303, USA}
\newcommand{\hiroshima}{Hiroshima University, Kagamiyama, Higashi-Hiroshima 739-8526, Japan}
\newcommand{\ihepprot}{IHEP Protvino, State Research Center of Russian Federation, Institute for High Energy Physics, Protvino, 142281, Russia}
\newcommand{\illuiuc}{University of Illinois at Urbana-Champaign, Urbana, IL 61801, USA}
\newcommand{\instpasczech}{Institute of Physics, Academy of Sciences of the Czech Republic, Na Slovance 2, 182 21 Prague 8, Czech Republic}
\newcommand{\isu}{Iowa State University, Ames, IA 50011, USA}
\newcommand{\jinrdubna}{Joint Institute for Nuclear Research, 141980 Dubna, Moscow Region, Russia}
\newcommand{\kek}{KEK, High Energy Accelerator Research Organization, Tsukuba, Ibaraki 305-0801, Japan}
\newcommand{\kfki}{KFKI Research Institute for Particle and Nuclear Physics of the Hungarian Academy of Sciences (MTA KFKI RMKI), H-1525 Budapest 114, POBox 49, Budapest, Hungary}
\newcommand{\korea}{Korea University, Seoul, 136-701, Korea}
\newcommand{\kurchatov}{Russian Research Center ``Kurchatov Institute", Moscow, Russia}
\newcommand{\kyoto}{Kyoto University, Kyoto 606-8502, Japan}
\newcommand{\labllr}{Laboratoire Leprince-Ringuet, Ecole Polytechnique, CNRS-IN2P3, Route de Saclay, F-91128, Palaiseau, France}
\newcommand{\lawllnl}{Lawrence Livermore National Laboratory, Livermore, CA 94550, USA}
\newcommand{\losalamos}{Los Alamos National Laboratory, Los Alamos, NM 87545, USA}
\newcommand{\lpc}{LPC, Universit{\'e} Blaise Pascal, CNRS-IN2P3, Clermont-Fd, 63177 Aubiere Cedex, France}
\newcommand{\lund}{Department of Physics, Lund University, Box 118, SE-221 00 Lund, Sweden}
\newcommand{\muenster}{Institut f\"ur Kernphysik, University of Muenster, D-48149 Muenster, Germany}
\newcommand{\myongji}{Myongji University, Yongin, Kyonggido 449-728, Korea}
\newcommand{\nagasaki}{Nagasaki Institute of Applied Science, Nagasaki-shi, Nagasaki 851-0193, Japan}
\newcommand{\newmex}{University of New Mexico, Albuquerque, NM 87131, USA }
\newcommand{\nmsu}{New Mexico State University, Las Cruces, NM 88003, USA}
\newcommand{\ornl}{Oak Ridge National Laboratory, Oak Ridge, TN 37831, USA}
\newcommand{\orsay}{IPN-Orsay, Universite Paris Sud, CNRS-IN2P3, BP1, F-91406, Orsay, France}
\newcommand{\peking}{Peking University, Beijing, People's Republic of China}
\newcommand{\pnpi}{PNPI, Petersburg Nuclear Physics Institute, Gatchina, Leningrad region, 188300, Russia}
\newcommand{\riken}{RIKEN, The Institute of Physical and Chemical Research, Wako, Saitama 351-0198, Japan}
\newcommand{\rikjrbrc}{RIKEN BNL Research Center, Brookhaven National Laboratory, Upton, NY 11973-5000, USA}
\newcommand{\rikkyo}{Physics Department, Rikkyo University, 3-34-1 Nishi-Ikebukuro, Toshima, Tokyo 171-8501, Japan}
\newcommand{\saispbstu}{Saint Petersburg State Polytechnic University, St. Petersburg, Russia}
\newcommand{\saopaulo}{Universidade de S{\~a}o Paulo, Instituto de F\'{\i}sica, Caixa Postal 66318, S{\~a}o Paulo CEP05315-970, Brazil}
\newcommand{\seoulnat}{System Electronics Laboratory, Seoul National University, Seoul, Korea}
\newcommand{\stonybrkc}{Chemistry Department, Stony Brook University, Stony Brook, SUNY, NY 11794-3400, USA}
\newcommand{\stonycrkp}{Department of Physics and Astronomy, Stony Brook University, SUNY, Stony Brook, NY 11794, USA}
\newcommand{\subatech}{SUBATECH (Ecole des Mines de Nantes, CNRS-IN2P3, Universit{\'e} de Nantes) BP 20722 - 44307, Nantes, France}
\newcommand{\tenn}{University of Tennessee, Knoxville, TN 37996, USA}
\newcommand{\titech}{Department of Physics, Tokyo Institute of Technology, Oh-okayama, Meguro, Tokyo 152-8551, Japan}
\newcommand{\tsukuba}{Institute of Physics, University of Tsukuba, Tsukuba, Ibaraki 305, Japan}
\newcommand{\vandy}{Vanderbilt University, Nashville, TN 37235, USA}
\newcommand{\waseda}{Waseda University, Advanced Research Institute for Science and Engineering, 17 Kikui-cho, Shinjuku-ku, Tokyo 162-0044, Japan}
\newcommand{\weizmann}{Weizmann Institute, Rehovot 76100, Israel}
\newcommand{\yonsei}{Yonsei University, IPAP, Seoul 120-749, Korea}
\affiliation{\abilene}
\affiliation{\banaras}
\affiliation{\bnl}
\affiliation{\caucr}
\affiliation{\charlesczech}
\affiliation{\ciae}
\affiliation{\cns}
\affiliation{\colorado}
\affiliation{\columbia}
\affiliation{\czechtech}
\affiliation{\dapnia}
\affiliation{\debrecen}
\affiliation{\elte}
\affiliation{\fit}
\affiliation{\fsu}
\affiliation{\gsu}
\affiliation{\hiroshima}
\affiliation{\ihepprot}
\affiliation{\illuiuc}
\affiliation{\instpasczech}
\affiliation{\isu}
\affiliation{\jinrdubna}
\affiliation{\kek}
\affiliation{\kfki}
\affiliation{\korea}
\affiliation{\kurchatov}
\affiliation{\kyoto}
\affiliation{\labllr}
\affiliation{\lawllnl}
\affiliation{\losalamos}
\affiliation{\lpc}
\affiliation{\lund}
\affiliation{\muenster}
\affiliation{\myongji}
\affiliation{\nagasaki}
\affiliation{\newmex}
\affiliation{\nmsu}
\affiliation{\ornl}
\affiliation{\orsay}
\affiliation{\peking}
\affiliation{\pnpi}
\affiliation{\riken}
\affiliation{\rikjrbrc}
\affiliation{\rikkyo}
\affiliation{\saispbstu}
\affiliation{\saopaulo}
\affiliation{\seoulnat}
\affiliation{\stonybrkc}
\affiliation{\stonycrkp}
\affiliation{\subatech}
\affiliation{\tenn}
\affiliation{\titech}
\affiliation{\tsukuba}
\affiliation{\vandy}
\affiliation{\waseda}
\affiliation{\weizmann}
\affiliation{\yonsei}
\author{A.~Adare}	\affiliation{\colorado}
\author{S.~Afanasiev}	\affiliation{\jinrdubna}
\author{C.~Aidala}	\affiliation{\columbia}
\author{N.N.~Ajitanand}	\affiliation{\stonybrkc}
\author{Y.~Akiba}	\affiliation{\riken} \affiliation{\rikjrbrc}
\author{H.~Al-Bataineh}	\affiliation{\nmsu}
\author{J.~Alexander}	\affiliation{\stonybrkc}
\author{K.~Aoki}	\affiliation{\kyoto} \affiliation{\riken}
\author{L.~Aphecetche}	\affiliation{\subatech}
\author{R.~Armendariz}	\affiliation{\nmsu}
% ...
\collaboration{PHENIX Collaboration} \noaffiliation

\begin{abstract}


Yields for \jpsi~production in \CuCu\ collisions at \sqrtsNN~=~200\,GeV 
have been measured 
over the rapidity range 
$|y| < 2.2$  and compared with results in \pp~and \AuAu~collisions at the same 
energy. The \CuCu\ data offer greatly improved precision over 
existing \AuAu\ data for \jpsi\ production in collisions with small 
to intermediate numbers of participants, in the range where the Quark Gluon Plasma 
transition threshold is predicted to lie. Cold nuclear matter estimates based on ad hoc 
fits to \dAu\ data describe the \CuCu\ data up to \Npart~$\sim$~50, corresponding 
to a Bjorken energy density of at least 1.5 GeV/fm$^3$.

\end{abstract}

\pacs{25.75.Dw,12.38.Mh,21.65.Qr,25.75.Nq}

\date{\today}

\maketitle



High-energy heavy-ion collisions provide the opportunity to study 
strongly interacting matter at very high energy densities where 
Quantum Chromodynamics (QCD) predicts a transition from normal 
nuclear matter to a de-confined system of quarks and gluons, the 
Quark-Gluon Plasma (QGP)~\cite{Harris:1996zx}.  At the 
Relativistic Heavy Ion Collider (RHIC) the 
energy density in central \AuAu\ collisions is well in excess of 
the critical energy density expected for this 
transition~\cite{Adcox:2004mh}.

Over the past twenty years, there has been intense theoretical and 
experimental work on \jpsi\ production.  First predicted by Matsui 
and Satz \cite{Matsui:1986dk}, suppression of quarkonia production 
in ultra-relativistic heavy ion collisions was expected to be an 
unambiguous signature for the formation of a QGP. It is now 
recognized that in order to interpret $J/\psi$ production as a QGP 
probe one has to consider cold nuclear matter effects such as 
initial state energy loss~\cite{Johnson:2000ph} and 
shadowing~\cite{Guzey:2004zp}, as well as charm quark energy 
loss~\cite{Baier:2000mf}, co-mover 
interactions~\cite{Gavin:1996fe}, corrections for feed-down from 
higher mass charmonium states, and secondary production mechanisms, 
such as recombination of initially uncorrelated $c\bar{c}$ 
pairs~\cite{Thews:2005vj}.

Experiment NA50 reported suppression of \jpsi\ production 
in \PbPb\ collisions at \sqrtsNN\ = 
17.3~GeV~\cite{Alessandro:2005ni} that exceeds expectations based 
on their measurements of cold nuclear matter effects in \pA\ 
collisions~\cite{Alessandro:2006ni}.  NA60 observed similar 
behavior in \InIn\ collisions at the same energy~\cite{Arnaldi:2007ni}.
The PHENIX experiment~\cite{nim_phenix} at RHIC has 
characterized effects of the nuclear medium on \jpsi\ 
production at \sqrtsNN\ = 200~GeV.  The basic invariant yield 
reference is obtained from \pp\ data~\cite{Adler:2003qs, 
Adler:2005ph, Adare:2006kf}.  Cold nuclear matter effects are 
studied using \dAu\ data~\cite{Adler:2005ph,Adare:ppg078}.  Cold and 
hot nuclear matter effects are studied for large numbers of 
participants (\Npart) using \AuAu\ 
data~\cite{Adler:2003rc,Adare:2006ns}, and for smaller \Npart\ 
using \CuCu\ data, the subject of this paper.  The results are 
presented as a nuclear modification factor, \RAA, the ratio of the 
yield in heavy ion collisions to the yield in \pp\ collisions 
scaled by the number of binary nucleon-nucleon collisions (\Ncoll), which is 
appropriate for point-like processes.

Lattice QCD calculations~\cite{karsch:2002} indicate
that the threshold energy density for QGP formation is of order 1
GeV/fm$^3$. At \sqrtsNN\ = 200 GeV this is expected to occur below \Npart\
$= 100$~\cite{phenix:PPG019}, in a region where \AuAu\ data have limited
statistical and systematic precision~\cite{Adare:2006ns}. High statistics 
measurements with the 
intermediate sized system \CuCu\ provide crucial information in that 
important region.

In this Letter we present results obtained by PHENIX during the 
2005 RHIC run on the production of \jpsi\ in \CuCu\ collisions at 
\sqrtsNN\ = 200 GeV. $\jpsi$ invariant yields were studied via 
$\jpsi \rightarrow e^+e^-$ decays measured at midrapidity with the 
central arm spectrometers ($|y| \leq 0.35$, $\Delta\phi = 2 \times 
90^{\circ}$), and $\jpsi \rightarrow \mu^+\mu^-$ decays measured at 
forward rapidity with the two muon arm spectrometers ($1.2 < |y| 
<2.2$, $\Delta\phi = 360^{\circ}$). Event centrality and the 
location of the collision vertex along the beam axis $(z_{vtx})$ 
are measured with two Beam-Beam Counters (BBC) located at 3.0 
$<|\eta|<$3.9. A Glauber model and a simulation of the BBC response 
was used to determine \Npart\ and \Ncoll\ and their systematic 
uncertainties for different collision centrality 
ranges~\cite{Adare:PPG084}.

Data were recorded using lepton triggers in coincidence with a 
minimum bias trigger which required a coincidence between the BBC 
detectors and a valid $z_{vtx}$.  After applying a cut of 
$|z_{vtx}| < 30$\,cm and quality assurance criteria, the data 
correspond to a sampled luminosity of about 
$2.1$ nb$^{-1}$ ($1.3$ nb$^{-1}$) in the $e^+e^-$ ($\mu^+\mu^-$) 
analysis.



Electron detection at midrapidity used the Drift Chambers for 
momentum measurement, the Pad Chambers for pattern recognition and 
track location, and the Ring Imaging Cherenkov (RICH) detector plus 
Electromagnetic Calorimeter (EMCal) for electron identification. 
Charged particle tracks were matched with a RICH ring and an EMCal 
hit to select electron candidates by requiring at least two RICH 
phototube hits inside an annulus around the projected ring center, 
ring quality cuts, track/cluster position matching cuts at the 
EMCal, and a cut on the ratio of EMCal energy to track momentum, 
$E/p - \mean{E/p} > -2\sigma$.


The $\jpsi \rightarrow e^+e^-$ trigger required one signal above a 
certain energy threshold in the EMCal and a matching RICH hit. Two 
energy thresholds were used during the run, 1.1\,GeV and 0.8\,GeV, 
yielding average \jpsi\ trigger efficiencies of $\sim$ 65\% and 
82\%, respectively. The $\jpsi \rightarrow e^+e^-$ signal 
extraction method was very similar to the method used in the recent 
\AuAu~\cite{Adare:2006ns} and \pp~\cite{Adare:2006kf} analyses. The 
like sign invariant mass spectrum was subtracted from the unlike 
sign spectrum. The remaining yield in the $\jpsi$ mass region ($2.9 
\leq M_{inv} \leq 3.3$\,GeV/$c^2$) was corrected for pairs lost to 
the radiative tail and pairs added by the continuum signal under 
the peak~\cite{Adare:2006kf}. The total $\jpsi$ count in the 
$e^+e^-$ channel was $\approx$ 2,050. The signal to background 
ratio $(S/B)$ was $\approx 1(6)$ for the most central(peripheral) 
collisions.


Muon detection at forward and backward rapidities used the muon 
arms, consisting of cathode strip tracking chambers in a magnetic 
field (MuTr) and Iarocci tube planes interleaved with thick steel 
absorbers (MuID). Muon candidates were identified by penetration to 
the last MuID gap, and their momenta were measured by their bend 
through the MuTr.


The dimuon trigger required two candidate tracks to penetrate the 
MuID, point back to the event vertex, and pass an opening angle cut 
($\theta\ >\ 19^\circ$). The dimuon combinatorial background was estimated 
using the product of the like sign counts, $2\sqrt{N^{++} \cdot N^{--}}$, 
and was subtracted from the unlike sign spectra.  The residual background 
(notably from the open charm pairs and Drell-Yan) was evaluated using an 
exponential form. The $\jpsi \rightarrow \mu^+\mu^-$ signal was estimated
by direct counting of the remaining pairs above the exponential fit in
the mass range $2.6 \leq M_{inv} \leq 3.6$\,GeV/$c^2$ and also by
using two fits with different parameterizations (single and double 
Gaussian) of the $\jpsi$ line shapes, as described 
in~\cite{Adare:2006kf,Adare:2006ns}. The average of the results gave 
the signal count and the variation gave the systematic error. The 
total $\jpsi$ yield was $\approx$ 9,000. The $S/B$ was $\approx 
0.3(1.0)$ for the most central(peripheral) collisions.


The $\jpsi$ invariant yield in the appropriate centrality, rapidity 
and transverse momentum bin is given by~:
\begin{equation}
\label{eq:xsection}
\frac{B_{ll}}{2\pi p_{\rm T}}\frac{d^2N_{J/\psi}}{dp_{\rm T} dy} = 
\frac{1}{2\pi p_{\rm T}}
\frac{N_{J/\psi}}{N_{\rm evt}\Delta y\Delta p_{\rm T} A\varepsilon},
\end{equation}
with $B_{ll}$ the branching ratio for $\jpsi\rightarrow l^+l^-$; 
$N_{J/\psi}$ the number of observed $\jpsi$; $N_{\rm evt}$ the 
number of events; $\Delta y$ the rapidity range; $\Delta p_{\rm T}$ 
the transverse momentum range, and $A\varepsilon$ the acceptance 
and efficiency correction (including trigger efficiency).


The determination of $A\varepsilon$ is done with a full GEANT 
simulation. The method is described in more detail 
in~\cite{Adare:2006kf}.  $A\varepsilon$ decreases with the 
collision centrality due to overlapping hits in the RICH and the 
EMCal in the central arm, and in the MuTr for the forward arms, 
leading to an increasing fraction of misreconstructed tracks in 
higher multiplicity events. This effect is evaluated by embedding 
simulated single $\jpsi$ events in real events. The efficiency loss 
in the most central collisions is 3\% for dielectron measurements 
and 20\% (16\%) for dimuon measurements at positive (negative) 
rapidity.


%

Systematic uncertainties in the measured $\jpsi$ invariant yield 
depend on $\jpsi$ rapidity and transverse momentum as well as on 
event centrality. Systematic uncertainties are grouped into three 
categories: point to point uncorrelated uncertainties (type A), 
which can move the points independently of each other, point to 
point correlated uncertainties (type B), which can move the points 
coherently, though not necessarily by the same amount, and global 
systematic uncertainties (type C). In all plots point to point 
uncorrelated systematic uncertainties and statistical uncertainties 
are quadratically summed and represented by vertical bars, point to 
point correlated systematic uncertainties are represented by boxes, 
and global systematic uncertainties (if any) are quoted.

%
\begin{table}[htb]
\caption{\label{tab:syst_error} Systematic error sources, values and
types for \RAA\ vs \Npart\ in the two rapidity intervals. Where a range is given, it is from peripheral 
to central collisions.}
\begin{ruledtabular} \begin{tabular}{rccc}
source&$|y|<0.3$&$|y|\in[1.2,2.2]$&type\\ \hline
signal extraction&6~\%&5-6~\%&A\\
detector + trigger efficiency& 1.4-5~\%~&3~\%&B\\
run by run variation&5~\%&2~\%&B\\
input $y$ + $p_{\rm T}$ distributions& 2~\% &3~\%&B\\
\Ncoll & 14-11~\% &  14-11~\% & B \\
\end{tabular} \end{ruledtabular} 
\end{table}

%
\begin{figure}[hbt]
\includegraphics[width=1.0\linewidth]{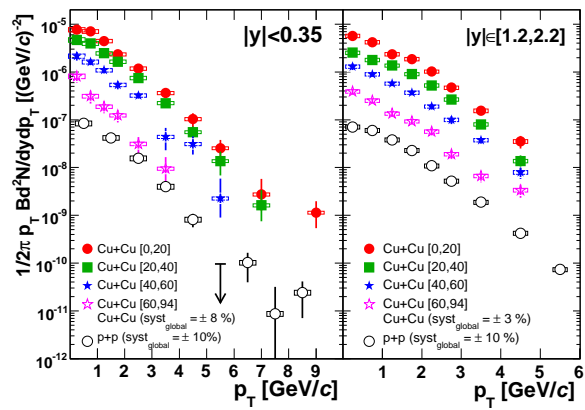}
\caption{(color online) $\jpsi$ yield vs \pT\ at mid (left) and forward (right) 
rapidity for different \CuCu\ centrality bins and for \pp~\cite{Adare:2006kf}. Uncertainties 
are described in the text.}
\label{fig:yield_pt}
\end{figure}

Systematic uncertainties of type A and B for \RAA\ vs \Npart\ are 
summarized in \tab{tab:syst_error}. Some uncertainties in the 
invariant yield, such as that on the acceptance, cancel out for 
\RAA\ and are not shown. Global systematic uncertainties for \RAA\ 
vs \Npart\ include the \pp~$\jpsi$ yield uncertainty and some \pp\ 
systematic errors that do not cancel when forming \RAA.

%
\begin{figure}[thb]
\includegraphics[width=1.0\linewidth]{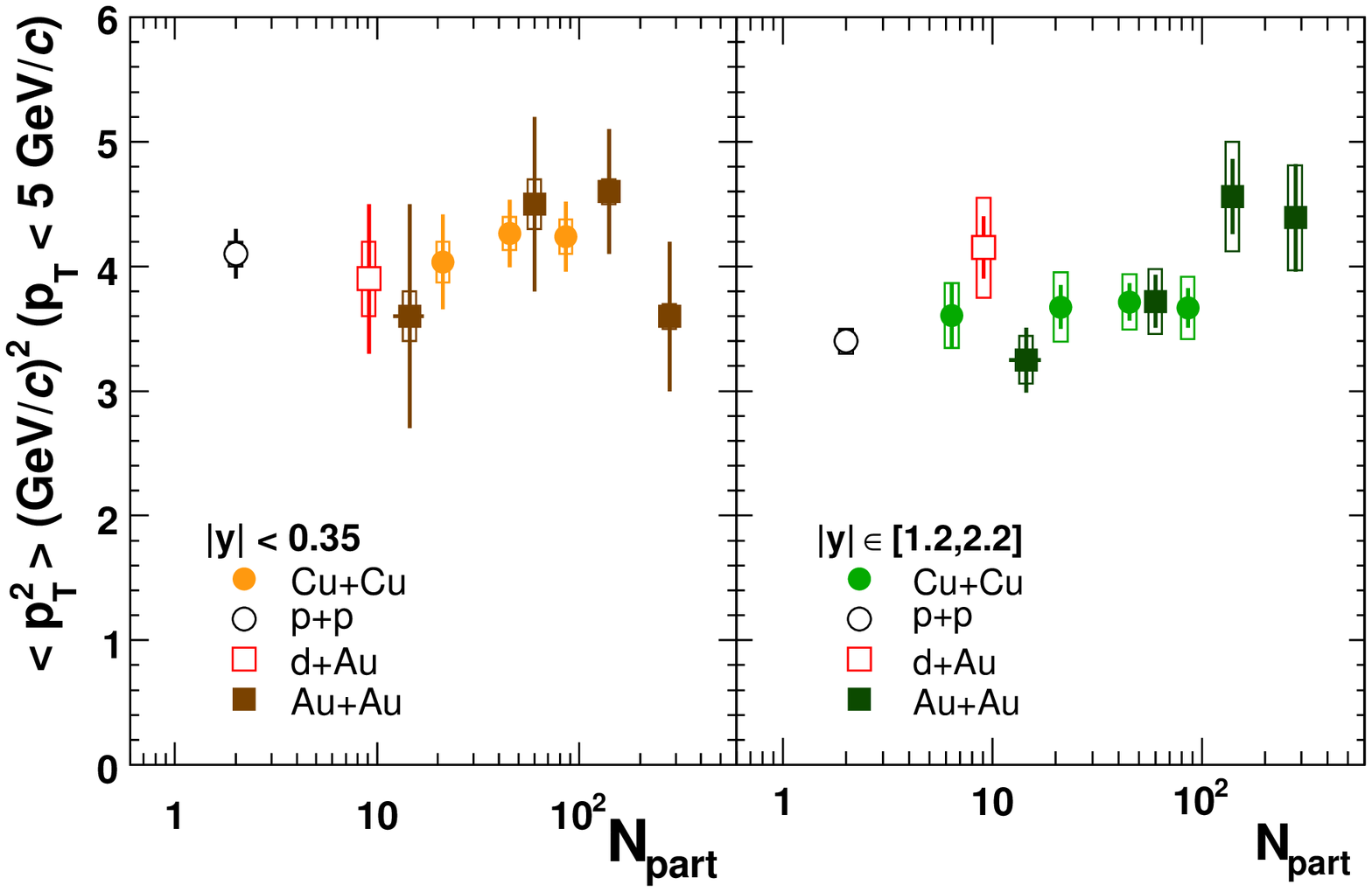}
\caption{(color online) The $\mean{p_{\rm T}^2}$~vs \Npart\ for 
$J/\psi$ production
 in \CuCu, \pp~\cite{Adare:2006kf}, \dAu~\cite{Adare:ppg078} and \AuAu~\cite{Adare:2006ns} 
 collisions at mid (left) and forward (right) rapidity.}
\label{fig:pt2}
\end{figure}

%
 \begin{figure}[thb]
 \includegraphics[width=1.0\linewidth]{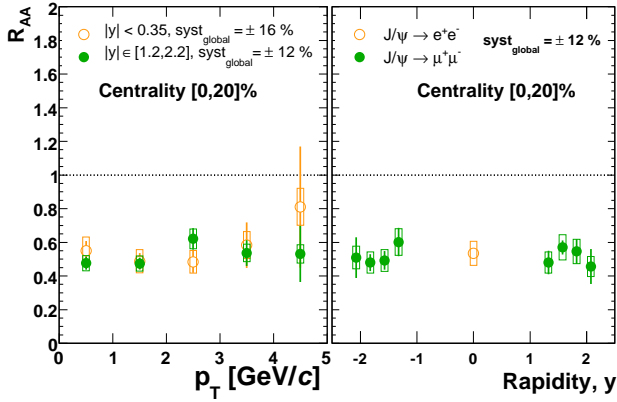}
 \caption{(color online) \RAA~vs~\pT\ (left) and $y$ (right) for $\jpsi$ production in the most central
 \CuCu\ collisions.}
 \label{fig:r_aa_pt_y}
 \end{figure}

Results for the two muon arms agree within uncertainties and are 
combined where appropriate. Fig.~\ref{fig:yield_pt} shows the 
$\jpsi$ yield vs \pT\ for different \CuCu\ centrality classes at 
mid and forward rapidity. As was done previously for the \AuAu\ 
case~\cite{Adare:2006ns}, the mean square transverse momentum, 
$\mean{p_{\rm T}^2}$, was calculated numerically from the data for 
\pT\ $< 5$ GeV/$c$. The \CuCu\ data are plotted vs \Npart\ and 
compared with the corresponding values from 
\AuAu~\cite{Adare:2006ns}, \dAu~\cite{Adare:ppg078} and 
\pp~\cite{Adare:2006kf} collisions in \fig{fig:pt2}. Within 
uncertainties, the data for \CuCu\ and \AuAu\ agree where they 
overlap in \Npart, and the $\mean{p_{\rm T}^2}$ for the \CuCu\ data 
seems independent of \Npart.

The $\raa$ values vs \pT\ and rapidity are shown in 
\fig{fig:r_aa_pt_y} for the 0--20\% most central \CuCu\ collisions.  
We see similar behavior for mid and forward rapidity, and there 
appears to be no \pT\ dependence in all centrality classes. The 
$\rm{RMS}$ width of the rapidity distribution (evaluated directly 
from the data) is identical, within $\sim 2-3\%$ uncertainties, in 
\pp\ collisions and in all centrality classes for \CuCu\ 
collisions.

%
 \begin{figure} [thb]
\includegraphics[width=1.0\linewidth]{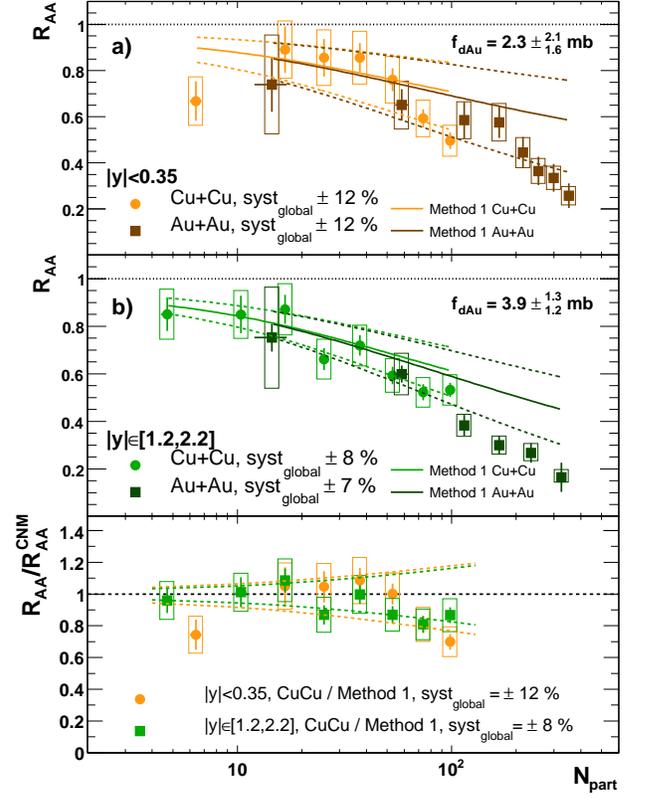}
 \caption{(color online) (a,b) \RAA\ vs \Npart\ for $\jpsi$ production
 in \CuCu\ and \AuAu~\cite{Adare:2006ns} collisions.  The curves are predictions 
 from ad hoc fits to \dAu\ data~\cite{Adare:ppg078} and are discussed in the text.  
 (c) Ratios of the measured \RAA\ values to the predicted cold nuclear matter \RAA.
 The dashed lines show the 1 $\sigma$ uncertainties from the \dAu\ fits.}

\label{fig:r_aa_n_part} \end{figure}

Figures~\ref{fig:r_aa_n_part}(a) and~\ref{fig:r_aa_n_part}(b) show similar behavior within
uncertainties for \RAA\ in \CuCu\ and \AuAu~\cite{Adare:2006ns}
collisions at comparable values of \Npart.
Theoretical calculations~\cite{Vogt:2004dh} including only modified 
initial parton distribution functions and an added $\jpsi-N$ 
breakup cross section were fitted in~\cite{Adare:ppg078} to \dAu\ 
$J/\psi$ \RAA\ data. 
The EKS98~\cite{Eskola:1998,Eskola:1999} and nDSg~\cite{deFlorian:2003qf} 
shadowing models were used. The fit was made simultaneously to all 
rapidities by optimizing the breakup cross section. 
While consistent with the low statistics \dAu\ data~\cite{Adare:ppg078}, 
this method leads to a model dependence of the CNM effects, since the 
rapidity shape is determined entirely by the shadowing model.
In an attempt to reduce this model dependence, we used 
a data-driven ad hoc model to parameterize the \dAu\ 
data~\cite{Adare:ppg078}. The ad hoc model uses EKS98 (method 1) 
and nDSg (method 2) shadowing parameterizations for the relative 
rapidity dependence within the fitted rapidity ranges, but the 
breakup cross section is replaced with a quantity, which we call 
$f$, that is optimized separately for $y$=0 and $|y|$=1.7. 
The fits using method 1 yielded $f_{dAu}=2.3 \pm^{2.1}_{1.6}$ mb at $y$=0 and $3.9 
\pm^{1.3}_{1.2}$ mb at $|y|$=1.7. The method 2 fits 
yielded $f_{dAu}=0.9 \pm^{1.9}_{1.8}$ mb at $y$=0 and $3.3 \pm^{1.3}_{1.2}$ 
mb at $|y|$=1.7. 
The resulting separate parameterizations of the \dAu\ data vs \Ncoll\ at mid and 
forward/backward rapidity can be projected to \CuCu\ and \AuAu\ using the corresponding parton 
distribution functions for Cu and Au~\cite{Vogt:2004dh}.
The results for method 1 are shown in \fig{fig:r_aa_n_part} as cold nuclear 
matter baseline \RAA\ curves calculated from the best fit values of 
$f$ (solid lines) and the one standard deviation uncertainty in $f$ 
(dashed lines). The method 2 heavy ion calculations are similar to 
those from method 1, leading to very similar conclusions, and are not shown 
in \fig{fig:r_aa_n_part}. 
In \fig{fig:r_aa_n_part}(c) the 
measured \RAA\ values
for \CuCu\ are shown divided by the method 1 calculations for \CuCu.
The \CuCu\ \RAA\ is seen to be consistent with the cold nuclear matter
projection within about 15\% uncertainties up to \Npart\ $\sim$ 50.
Given the uncertainty in the cold nuclear matter reference at larger 
\Npart\ values, we can not currently draw any strong conclusions there. 
However PHENIX completed in February 2008 a second \dAu\ run, 
with approximately 30 times the statistics of the first \dAu\ run in 2003.
With the new reference \dAu\ data, we expect to be able 
to identify if and where the measured \CuCu\ \RAA\ departs from the cold 
nuclear matter baseline.


In summary, we present high statistics \jpsi\ data from \CuCu\ 
collisions at RHIC, providing for the first time detailed 
information on \RAA\ and $\mean{p_{\rm T}^2}$ for \Npart\ $< 100$.  
The RMS values of the rapidity distributions at all centralities 
are consistent with that for \pp, and the measured $\mean{p_{\rm 
T}^2}$ for \pT\ $< 5$ GeV/$c$ is nearly independent of centrality 
and rapidity. At similar values of \Npart, \RAA\ and $\mean{p_{\rm 
T}^2}$ are found to agree within errors for \CuCu\ and \AuAu\ collisions. 
Cold nuclear matter calculations based on ad hoc fits to \dAu\ data
reproduce the peripheral \CuCu\ data well up to \Npart\ $\sim$ 50,
corresponding to $\epsilon_{Bjorken}~\tau~\sim$ 1.5 GeV/fm$^2/c$~\cite{phenix:PPG019}, 
where $\epsilon_{Bjorken}$ is the Bjorken energy density 
and $\tau$ is the formation time. For an estimate of the thermalized
energy density, hydrodynamical models give thermalization times in the
range of 0.6 fm/$c$ to 1.0 fm/$c$~\cite{Adcox:2004mh}, which implies that
cold nuclear matter effects dominate \jpsi\ production up to
thermalized energy densities of $\sim$ 1.5 to 2.5 GeV/fm$^3$.


We thank the staff of the Collider-Accelerator and 
Physics Departments at BNL for their vital contributions.  
We acknowledge support from 
the Office of Nuclear Physics in DOE Office of Science and NSF (U.S.A.), 
MEXT and JSPS (Japan), 
CNPq and FAPESP (Brazil), 
NSFC (China), 
MSMT (Czech Republic),
IN2P3/CNRS, and CEA (France), 
BMBF, DAAD, and AvH (Germany), 
OTKA (Hungary), 
DAE (India), 
ISF (Israel), 
KRF and KOSEF (Korea), 
MES, RAS, and FAAE (Russia),
VR and KAW (Sweden), 
U.S. CRDF for the FSU, 
US-Hungarian NSF-OTKA-MTA, 
and US-Israel BSF.



\begin{thebibliography}{22}
\expandafter\ifx\csname natexlab\endcsname\relax\def\natexlab#1{#1}\fi
\expandafter\ifx\csname bibnamefont\endcsname\relax
  \def\bibnamefont#1{#1}\fi
\expandafter\ifx\csname bibfnamefont\endcsname\relax
  \def\bibfnamefont#1{#1}\fi
\expandafter\ifx\csname citenamefont\endcsname\relax
  \def\citenamefont#1{#1}\fi
\expandafter\ifx\csname url\endcsname\relax
  \def\url#1{\texttt{#1}}\fi
\expandafter\ifx\csname urlprefix\endcsname\relax\def\urlprefix{URL }\fi
\providecommand{\bibinfo}[2]{#2}
\providecommand{\eprint}[2][]{\url{#2}}

\bibitem[{\citenamefont{Harris and {M\"uller}}(1996)}]{Harris:1996zx}
\bibinfo{author}{\bibfnamefont{J.~W.} \bibnamefont{Harris}} \bibnamefont{and}
  \bibinfo{author}{\bibfnamefont{B.}~\bibnamefont{{M\"uller}}},
  \bibinfo{journal}{Ann. Rev. Nucl. Part. Sci.} \textbf{\bibinfo{volume}{46}},
  \bibinfo{pages}{71} (\bibinfo{year}{1996}).

\bibitem[{\citenamefont{Adcox et~al.}(2005)}]{Adcox:2004mh}
\bibinfo{author}{\bibfnamefont{K.}~\bibnamefont{Adcox}} \bibnamefont{et~al.},
  \bibinfo{journal}{Nucl. Phys.} \textbf{\bibinfo{volume}{A757}},
  \bibinfo{pages}{184} (\bibinfo{year}{2005}).

\bibitem[{\citenamefont{Matsui and Satz}(1986)}]{Matsui:1986dk}
\bibinfo{author}{\bibfnamefont{T.}~\bibnamefont{Matsui}} \bibnamefont{and}
  \bibinfo{author}{\bibfnamefont{H.}~\bibnamefont{Satz}},
  \bibinfo{journal}{Phys. Lett.} \textbf{\bibinfo{volume}{B178}},
  \bibinfo{pages}{416} (\bibinfo{year}{1986}).

\bibitem[{\citenamefont{Johnson et~al.}(2001)}]{Johnson:2000ph}
\bibinfo{author}{\bibfnamefont{M.~B.} \bibnamefont{Johnson}}
  \bibnamefont{et~al.}, \bibinfo{journal}{Phys. Rev. Lett.}
  \textbf{\bibinfo{volume}{86}}, \bibinfo{pages}{4483} (\bibinfo{year}{2001}).

\bibitem[{\citenamefont{Guzey et~al.}(2004)}]{Guzey:2004zp}
\bibinfo{author}{\bibfnamefont{V.}~\bibnamefont{Guzey}} \bibnamefont{et~al.}, 
  \bibinfo{journal}{Phys. Lett.}
  \textbf{\bibinfo{volume}{B603}}, \bibinfo{pages}{173} (\bibinfo{year}{2004}).

\bibitem[{\citenamefont{Baier et~al.}(2000)}]{Baier:2000mf}
\bibinfo{author}{\bibfnamefont{R.}~\bibnamefont{Baier}} \bibnamefont{et~al.},
  \bibinfo{journal}{Ann. Rev. Nucl. Part. Sci.} \textbf{\bibinfo{volume}{50}},
  \bibinfo{pages}{37} (\bibinfo{year}{2000}).

\bibitem[{\citenamefont{Gavin and Vogt}(1996)}]{Gavin:1996fe}
\bibinfo{author}{\bibfnamefont{S.}~\bibnamefont{Gavin}} \bibnamefont{and}
  \bibinfo{author}{\bibfnamefont{R.}~\bibnamefont{Vogt}},
  \bibinfo{journal}{Nucl. Phys.} \textbf{\bibinfo{volume}{A610}},
  \bibinfo{pages}{442c} (\bibinfo{year}{1996}).

\bibitem[{\citenamefont{Thews and Mangano}(2006)}]{Thews:2005vj}
\bibinfo{author}{\bibfnamefont{R.~L.} \bibnamefont{Thews}} \bibnamefont{and}
  \bibinfo{author}{\bibfnamefont{M.~L.} \bibnamefont{Mangano}},
  \bibinfo{journal}{Phys. Rev.} \textbf{\bibinfo{volume}{C73}},
  \bibinfo{pages}{014904} (\bibinfo{year}{2006}).

\bibitem[{\citenamefont{Alessandro et~al.}(2005)}]{Alessandro:2005ni}
\bibinfo{author}{\bibfnamefont{B.}~\bibnamefont{Alessandro}}
  \bibnamefont{et~al.}, \bibinfo{journal}{Eur. Phys. J.}
  \textbf{\bibinfo{volume}{C39}}, \bibinfo{pages}{335} (\bibinfo{year}{2005}).

\bibitem[{\citenamefont{Alessandro et~al.}(2007)}]{Alessandro:2006ni}
\bibinfo{author}{\bibfnamefont{B.}~\bibnamefont{Alessandro}}
  \bibnamefont{et~al.}, \bibinfo{journal}{Eur. Phys. J.}
  \textbf{\bibinfo{volume}{C48}}, \bibinfo{pages}{329} (\bibinfo{year}{2007}).

\bibitem[{\citenamefont{Arnaldi et~al.}(2007)}]{Arnaldi:2007ni}
\bibinfo{author}{\bibfnamefont{R.}~\bibnamefont{Arnaldi}} \bibnamefont{et~al.},
\eprint{arXiv:0706.4361 [nucl-ex].}

\bibitem[{\citenamefont{Adcox et~al.}(2003)}]{nim_phenix}
\bibinfo{author}{\bibfnamefont{K.}~\bibnamefont{Adcox}} \bibnamefont{et~al.},
  \bibinfo{journal}{Nucl. Instrum. Meth.} \textbf{\bibinfo{volume}{A499}},
  \bibinfo{pages}{469} (\bibinfo{year}{2003}).

\bibitem[{\citenamefont{Adler et~al.}(2004{\natexlab{a}})}]{Adler:2003qs}
\bibinfo{author}{\bibfnamefont{S.~S.} \bibnamefont{Adler}}
  \bibnamefont{et~al.}, \bibinfo{journal}{Phys. Rev. Lett.}
  \textbf{\bibinfo{volume}{92}}, \bibinfo{pages}{051802}
  (\bibinfo{year}{2004}{\natexlab{a}}).

\bibitem[{\citenamefont{Adler et~al.}(2006)}]{Adler:2005ph}
\bibinfo{author}{\bibfnamefont{S.~S.} \bibnamefont{Adler}}
  \bibnamefont{et~al.}, \bibinfo{journal}{Phys. Rev. Lett.}
  \textbf{\bibinfo{volume}{96}}, \bibinfo{pages}{012304}
  (\bibinfo{year}{2006}).

\bibitem[{\citenamefont{Adare et~al.}(2007{\natexlab{a}})}]{Adare:2006kf}
\bibinfo{author}{\bibfnamefont{A.}~\bibnamefont{Adare}} \bibnamefont{et~al.},
  \bibinfo{journal}{Phys. Rev. Lett.} \textbf{\bibinfo{volume}{98}},
  \bibinfo{pages}{232002} (\bibinfo{year}{2007}{\natexlab{a}}).

\bibitem[{\citenamefont{Adare et~al.}(2008{\natexlab{b}})}]{Adare:ppg078}
\bibinfo{author}{\bibfnamefont{A.}~\bibnamefont{Adare}} \bibnamefont{et~al.},
\bibinfo{journal}{Phys. Rev.}
  \textbf{\bibinfo{volume}{C77}}, \bibinfo{pages}{024912}
  (\bibinfo{year}{2008}{\natexlab{b}}).

\bibitem[{\citenamefont{Adler et~al.}(2004{\natexlab{b}})}]{Adler:2003rc}
\bibinfo{author}{\bibfnamefont{S.~S.} \bibnamefont{Adler}}
  \bibnamefont{et~al.}, \bibinfo{journal}{Phys. Rev.}
  \textbf{\bibinfo{volume}{C69}}, \bibinfo{pages}{014901}
  (\bibinfo{year}{2004}{\natexlab{b}}).

\bibitem[{\citenamefont{Adare et~al.}(2007{\natexlab{c}})}]{Adare:2006ns}
\bibinfo{author}{\bibfnamefont{A.}~\bibnamefont{Adare}} \bibnamefont{et~al.},
  \bibinfo{journal}{Phys. Rev. Lett.} \textbf{\bibinfo{volume}{98}},
  \bibinfo{pages}{232301} (\bibinfo{year}{2007}{\natexlab{c}}).

\bibitem[{\citenamefont{Karsch}(2002)}]{karsch:2002}
\bibinfo{author}{\bibfnamefont{F.}~\bibnamefont{Karsch}}, \bibinfo{journal}{Lect. 
  Notes Phys.} \textbf{\bibinfo{volume}{583}}, \bibinfo{pages}{209}
  (\bibinfo{year}{2002}).

\bibitem[{\citenamefont{Adler et~al.}(2005)}]{phenix:PPG019}
\bibinfo{author}{\bibfnamefont{S.~S.} \bibnamefont{Adler}}
  \bibnamefont{et~al.}, \bibinfo{journal}{Phys. Rev.}
  \textbf{\bibinfo{volume}{C71}}, \bibinfo{pages}{034908}
  (\bibinfo{year}{2005}).

\bibitem[{\citenamefont{Adare et~al.}(200X)}]{Adare:PPG084}
\bibinfo{author}{\bibfnamefont{A.}~\bibnamefont{Adare}} \bibnamefont{et~al.},
\eprint{arXiv:0801.4555 [nucl-ex]}.

\bibitem[{\citenamefont{Vogt}(2005)}]{Vogt:2004dh}
\bibinfo{author}{\bibfnamefont{R.}~\bibnamefont{Vogt}}, \bibinfo{journal}{Phys.
  Rev.} \textbf{\bibinfo{volume}{C71}}, \bibinfo{pages}{054902}
  (\bibinfo{year}{2005}).

\bibitem[{\citenamefont{Eskola et~al.}(1998)}]{Eskola:1998}
\bibinfo{author}{\bibfnamefont{K.~J.} \bibnamefont{Eskola}}
  \bibnamefont{et~al.}, \bibinfo{journal}{Nucl. Phys.}
  \textbf{\bibinfo{volume}{B535}}, \bibinfo{pages}{351} (\bibinfo{year}{1998}).

\bibitem[{\citenamefont{Eskola et~al.}(1990)}]{Eskola:1999}
\bibinfo{author}{\bibfnamefont{K.~J.} \bibnamefont{Eskola}}
  \bibnamefont{et~al.}, \bibinfo{journal}{Eur. Phys. J.}
  \textbf{\bibinfo{volume}{C9}}, \bibinfo{pages}{61} (\bibinfo{year}{1998}).

\bibitem[{\citenamefont{de~Florian and Sassot}(2004)}]{deFlorian:2003qf}
\bibinfo{author}{\bibfnamefont{D.}~\bibnamefont{de~Florian}} \bibnamefont{and}
  \bibinfo{author}{\bibfnamefont{R.}~\bibnamefont{Sassot}},
  \bibinfo{journal}{Phys. Rev.} \textbf{\bibinfo{volume}{D69}},
  \bibinfo{pages}{074028} (\bibinfo{year}{2004}).

\end{thebibliography}

\end{document}